\documentclass[10pt,twoside]{article}
\usepackage[paperwidth=8.5in,paperheight=11in,left=1.55in,right=1.55in,top=1.45in,bottom=1.15in,headheight=15pt,headsep=0.28in]{geometry}
\usepackage{amsmath,amssymb,amsfonts}
\usepackage{newtxtext,newtxmath}
\usepackage{graphicx}
\usepackage[font=small,labelfont=bf,labelsep=period]{caption}
\usepackage[caption=false]{subfig}
\usepackage[numbers,super,sort&compress]{natbib}
\usepackage{titlesec}
\usepackage{fancyhdr}
\usepackage{lastpage}
\usepackage{microtype}
\usepackage{url}
\usepackage{hyperref}

\DeclareMathOperator{\sech}{sech}

\numberwithin{equation}{section}

\titleformat{\section}{\normalfont\bfseries\large}{\thesection.}{0.5em}{}
\titleformat{\subsection}{\normalfont\bfseries}{\thesubsection.}{0.5em}{}
\titleformat{\subsubsection}{\normalfont\itshape}{\thesubsubsection.}{0.5em}{}
\titlespacing*{\section}{0pt}{2.2ex plus 0.8ex minus 0.2ex}{1.0ex plus 0.2ex}
\titlespacing*{\subsection}{0pt}{1.8ex plus 0.6ex minus 0.2ex}{0.8ex plus 0.2ex}
\titlespacing*{\subsubsection}{0pt}{1.4ex plus 0.4ex minus 0.2ex}{0.6ex plus 0.2ex}

\newcommand{\journalname}{American Physical Society: Physical Review E}
\newcommand{\volumeinfo}{Vol. 113, Iss. 6 (2026) 064205-1--064205-12\\ https://link.aps.org/doi/10.1103/x5m6-wqkm}
\newcommand{\copyrightline}{\noindent{\footnotesize
		\textcopyright{} 2026 The Author(s).
		Published by the American Physical Society under the terms of the
		Creative Commons Attribution 4.0 International license.
}}
\newcommand{\shortauthors}{D. Ahmad et al.}
\newcommand{\shorttitle}{Supratransmission in Lattices with Purely Nonlinear Coupling}
\newcommand{\papertitle}{Supratransmission in lattices with purely nonlinear coupling}
\newcommand{\receiveddate}{\today}
\newcommand{\paperkeywords}{Supratransmission; purely nonlinear coupling; discrete $p$-Schr\"odinger equation; evanescent waves; saddle-node bifurcation.}
\newcommand{\paperabstract}{%
	Supratransmission is examined in nonlinear lattices with purely nonlinear coupling, extending the phenomenon to systems that lack a linear pass band. 
	In contrast to standard lattices with mixed linear--nonlinear interactions, the present model has no linear spectrum, so energy propagation arises entirely from nonlinear effects. 
	Asymptotic analysis yields a discrete $p$-Schr\"odinger (DpS) equation that 
	{provides an accurate description in the weak- and intermediate-coupling regimes and offers qualitative insight in the strong-coupling regime}. 
	Perturbation provides analytical approximations for the critical driving amplitude, explicitly showing its dependence on the driving frequency, coupling strength, and the nonlinearity exponent $p$. 
	The analysis identifies a non-trivial dependence of the critical amplitude on $p$, with distinct trends in different coupling regimes. 
	Numerical continuation and direct simulations {validate the theory in regimes where the asymptotic reduction is applicable and show good agreement across a wide range of parameters}. 
	The results establish supratransmission in fully nonlinear lattices and clarify the associated energy-transport mechanisms, with relevance to mechanical lattices, tunable metamaterials, and nonlinear optical arrays.}

\fancypagestyle{firstpage}{%
	\fancyhf{}%
	\fancyfoot[C]{\small\thepage}%
}

\makeatletter
\newcommand{\wsmaketitle}{%
	\thispagestyle{firstpage}%
	\noindent\begin{minipage}[t]{0.98\textwidth}\raggedright\small
		\journalname\\
		\volumeinfo\\
		\copyrightline
	\end{minipage}\hfill
	\begin{center}
		{\bfseries\fontsize{13.5}{16}\selectfont\MakeUppercase{\papertitle}\par}
		\vspace{0.30in}
		{\normalsize D. AHMAD\par}
		{\itshape\small Department of Mathematics, Khalifa University,\par
			PO Box 127788, Abu Dhabi, United Arab Emirates\par
			On leave: Department of Mathematics, Universitas Negeri Padang, Padang 25131, Indonesia\par}
		\vspace{0.08in}
		{\normalsize T.-Y. KIM\par}
		{\itshape\small Department of Civil and Environmental Engineering, Khalifa University,\par
			PO Box 127788, Abu Dhabi, United Arab Emirates\par
			Advanced Digital \& Additive Manufacturing Group, Khalifa University,\par
			PO Box 127788, Abu Dhabi, United Arab Emirates\par}
		\vspace{0.08in}
		{\normalsize A. SCHIFFER\par}
		{\itshape\small Department of Mechanical and Nuclear Engineering, Khalifa University,\par
			PO Box 127788, Abu Dhabi, United Arab Emirates\par}
		\vspace{0.08in}
		{\normalsize J. YANG\par}
		{\itshape\small Department of Mechanical Engineering, Seoul National University,\par
			1 Gwanak-ro, Gwanak-gu, Seoul, 08826, Republic of Korea\par}
		\vspace{0.08in}
		{\normalsize H. SUSANTO\par}
		{\itshape\small Department of Mathematics, Khalifa University,\par
			PO Box 127788, Abu Dhabi, United Arab Emirates\par}
		\vspace{0.12in}
		{\small Received \receiveddate\par}
	\end{center}
	\vspace{0.16in}
	\begin{quote}\small\noindent\paperabstract\end{quote}
	\vspace{0.06in}
	\noindent{\small\textit{Keywords}: \paperkeywords\par}
	\vspace{0.28in}
}
\makeatother

\hypersetup{hidelinks,pdfauthor={D. Ahmad, T.-Y. Kim, A. Schiffer, J. Yang, H. Susanto},pdftitle={Supratransmission in lattices with purely nonlinear coupling}}

\begin{document}
	
	\wsmaketitle
	
	\section{Introduction}\label{sec1}
	
	Energy transport in nonlinear lattices is a central topic in the study of wave propagation in discrete media, with applications in mechanical metamaterials \cite{bertoldi2017flexible}, granular crystals \cite{nesterenko2013dynamics,sen2008solitary}, optical waveguide arrays \cite{christodoulides2003discretizing,kartashov2009soliton}, and biological systems \cite{peyrard2004nonlinear}. 
	Among the established models for such media, Fermi--Pasta--Ulam--Tsingou (FPUT) lattices play a key role. Initially introduced in the context of thermalization \cite{fermi1955studies,dauxois2008fermi}, these lattices now serve as standard frameworks for examining nonlinear wave propagation, energy localization, and coherent structures such as solitons and breathers \cite{campbell2005introduction,berman2005fermi,flach2008discrete}. Their flexibility has made them effective reduced models for a wide range of engineered and natural materials \cite{kevrekidis2009discrete}.
	
	The general FPUT-type Hamiltonian is
	\begin{equation}
		\mathcal{H} = \sum_n H_n 
		:= \sum_n \frac12 \dot{y}_n^2 + W(y_n) + V(y_{n+1}-y_n),
		\label{H}
	\end{equation}
	where $V$ is the interaction potential and $W$ is an onsite (substrate) potential. 
	The corresponding equations of motion are
	\begin{equation}
		\ddot{y}_n + W'(y_n) 
		= V'(y_{n+1}-y_n) + V'(y_{n-1}-y_n),
		\label{eom}
	\end{equation}
	capturing the interplay between local nonlinear restoring forces and nearest-neighbor coupling.
	
	A widely studied example is the Hertzian contact potential,
	\begin{equation}
		V(x) = \frac{2}{5} C \,[x]_+^{5/2},
		\label{Hertz}
	\end{equation}
	with $[a]_+ = \max(a,0)$, which arises in granular media \cite{nesterenko2013dynamics,nguyen2012shock}.  
	Linear and nonlinear (e.g., power-law) coupling laws can be recovered by choosing different interaction exponents, thereby modeling systems ranging from optical lattices to soft granular materials \cite{chong2018coherent}. The linear response of granular chains depends strongly on precompression: without it (the ``sonic vacuum'' regime), the linearized stiffness vanishes, and wave propagation is governed solely by nonlinear effects; with precompression, a finite sound speed is restored \cite{chong2018coherent,sen2008solitary}. 
	
	{To place the Hertzian interaction within a broader context, we introduce a general class of power-law interactions. In this framework, Eq.~\eqref{Hertz} serves as a representative example, while the simulations and analysis are based on the generic formulation introduced below.}
	
	The model \eqref{eom} combined with \eqref{Hertz} and a nonlinear onsite potential
	\begin{equation}
		W(x) = \tfrac12 x^2 + \mathcal{O}(x^3)
	\end{equation}
	describes, for example, Newton's cradle \cite{james2011nonlinear} or chains of beads mounted on elastic substrates \cite{starosvetsky2012strongly}.  
	In Newton's cradle, gravity introduces a linear restoring term together with higher-order corrections, yielding a nonlinear onsite response \cite{hutzler2004rocking}.  
	A quartic coupling of the form
	\begin{equation}
		V(x) = \frac14 C\, x^4,
		\label{quartic}
	\end{equation}
	together with a nonlinear onsite potential
	\begin{equation}
		W(x) = \frac12 x^2 - \frac14 \gamma x^4,
		\label{onsite_quartic}
	\end{equation}
	has been used to study compactons and compacton-like structures \cite{kivshar1994compactons,comte2002exact,rosenau2005compact,gorbach2005compactlike,christodoulidi2025energy}.  
	Experimental implementations of nonlinear coupling have been demonstrated in mechanical systems \cite{dusuel1998kinks}, and similar nonlinear dispersion arises in averaged models with fast temporal modulation \cite{abdullaev2010compactons,d2015multidimensional}.
	
	To unify the interaction laws discussed above, we consider the generic power-law interaction
	\begin{equation}
		V(x) = \frac{C}{p+1}\, |x|^{p+1},
		\label{Vgen}
	\end{equation}
	which includes the harmonic case for $p=1$ and the quartic interaction \eqref{quartic} for $p=3$. {The Hertzian case corresponds to the exponent $p=3/2$, with Eq.~\eqref{Hertz} obtained by imposing unilateral (tensionless) contact via the positive-part operator. Unlike the Hertzian case, the interaction \eqref{Vgen} is symmetric and smooth, facilitating an analytical treatment but leading to different contact dynamics. In the following, this formulation provides the basis for the interaction used in the simulations and subsequent analysis. }
	
	With the sinusoidal onsite potential $W(x)=1-\cos x$ (whose small-amplitude expansion reduces to \eqref{onsite_quartic} with $\gamma=1/6$), the governing equation becomes
	\begin{equation}
		\begin{aligned}
			\ddot{y}_n + \sin(y_n) 
			=&\, C\bigl[
			\operatorname{sign}(y_{n+1}-y_n)\,|y_{n+1}-y_n|^{p} \\
			&\,+\, \operatorname{sign}(y_{n-1}-y_n)\,|y_{n-1}-y_n|^{p}
			\bigr].
		\end{aligned}
		\label{gov}
	\end{equation}
	For $p=1$, this reduces to the standard discrete sine--Gordon equation.
	
	We focus on \emph{supratransmission}, a nonlinear mechanism enabling energy transport at driving frequencies lying outside the linear pass band \cite{zakharov2023effect}. 
	The boundary driving is prescribed as
	\begin{equation}
		y_0(t) = F \cos(\omega t),
		\label{drive}
	\end{equation}
	with amplitude $F$ and frequency $\omega$.
	
	The study of time-periodic responses in nonlinear lattices is naturally formulated within the nonlinear response manifold framework developed by Kopidakis and Aubry \cite{kopidakis2000discrete,kopidakis2000intraband}, and subsequently employed in the analysis of nonlinear transmission and self-induced transparency phenomena \cite{maniadis2006energy,johansson2009transmission}. In this approach, a harmonic forcing is applied at a lattice site, and the corresponding time-periodic solutions are continued with respect to the driving amplitude. These solutions define a nonlinear manifold in the space of lattice responses and driving parameters, whose geometric structure encodes bifurcations and resonant interactions between lattice nonlinearity and the linear phonon spectrum.

	When $p=1$, linearization of \eqref{gov} about $y_n = 0$ yields the dispersion relation
	\begin{equation}
		\omega^2 = 1 + 2C(1 - \cos k), \qquad k \in [0,2\pi],
		\label{disper}
	\end{equation}
	and the admissible frequency band
	\[
	|\omega| \in [1, \sqrt{1+4C}].
	\]
	Supratransmission for \(|\omega|<1\) has been reported experimentally and numerically in granular chains and related systems \cite{lydon2015nonlinear,cui2022interaction}. In the $p=1$ case, \eqref{gov} corresponds to the model of Geniet and Leon \cite{geniet2002energy}, and modifications involving nonlinear onsite terms have been shown to reduce the supratransmission threshold \cite{alima2017influence}.
	
	The studies above share a common feature: a linear spectrum arising from either linear coupling or precompression. In contrast, when $p \neq 1$, the coupling becomes \emph{purely nonlinear}, and the system admits no linear pass band. Energy transport is therefore mediated primarily by nonlinear interactions. This setting alters the onset criteria for supratransmission and introduces new dynamical mechanisms.
	
	In this work, we analyze supratransmission in lattices governed by purely nonlinear coupling potentials.  
	Using asymptotic analysis, we derive a discrete $p$-Schr\"odinger (DpS) equation that provides an accurate description in weak and intermediate coupling, while offering qualitative insight into the strong-coupling regime, where the approximation becomes less accurate.  
	The resulting approximation yields explicit expressions for the critical driving amplitude and reveals its dependence on the interaction exponent $p$.  
	Two distinct supratransmission mechanisms are identified within the DpS framework in different parameter regimes and appear to govern the dynamics of the full system: one associated with a fold bifurcation of stationary states and the other with instability along the lower stationary branch.  
	Numerical continuation and direct simulations validate these predictions and demonstrate strong agreement with the analytical approximations.
	
	The paper is organized as follows.  
	Section~\ref{sec2} presents numerical simulations of supratransmission in the full sine--Gordon model with nonlinear coupling.  
	Section~\ref{sec3} derives the DpS equation and discusses its range of validity.  
	Section~\ref{sec4} describes the computation and stability analysis of evanescent waves.  
	Section~\ref{sec5} compares numerical results for the original system and the DpS approximation.  
	Section~\ref{sec6} develops analytical formulas for the critical driving amplitude and compares them with numerics.  
	Section~\ref{sec7} summarizes the main findings and outlines perspectives for future work.
	
	\section{Numerical simulations}\label{sec2}
	
	\begin{figure}[tbhp!]
		\centering
		\subfloat[$C=5$, $f=1.83$.]{\includegraphics[width=0.48\linewidth]{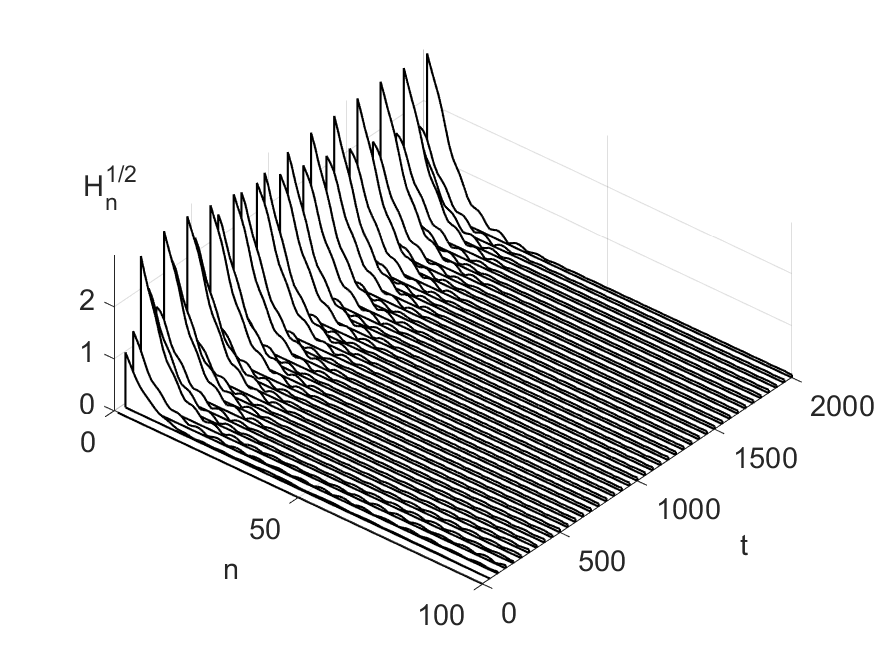}\label{subfig:supra_dynamics1_a}}
		\subfloat[$C=5$, $f=1.84$.]{\includegraphics[width=0.48\linewidth]{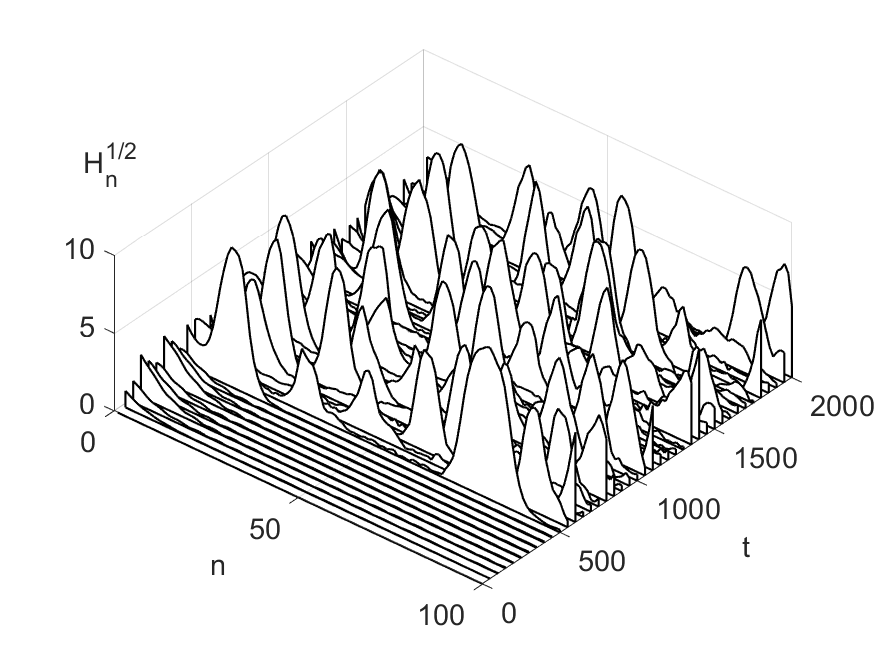}\label{subfig:supra_dynamics1_b}}\\
		\subfloat[$C=0.1$, $f=2.41$.]{\includegraphics[width=0.48\linewidth]{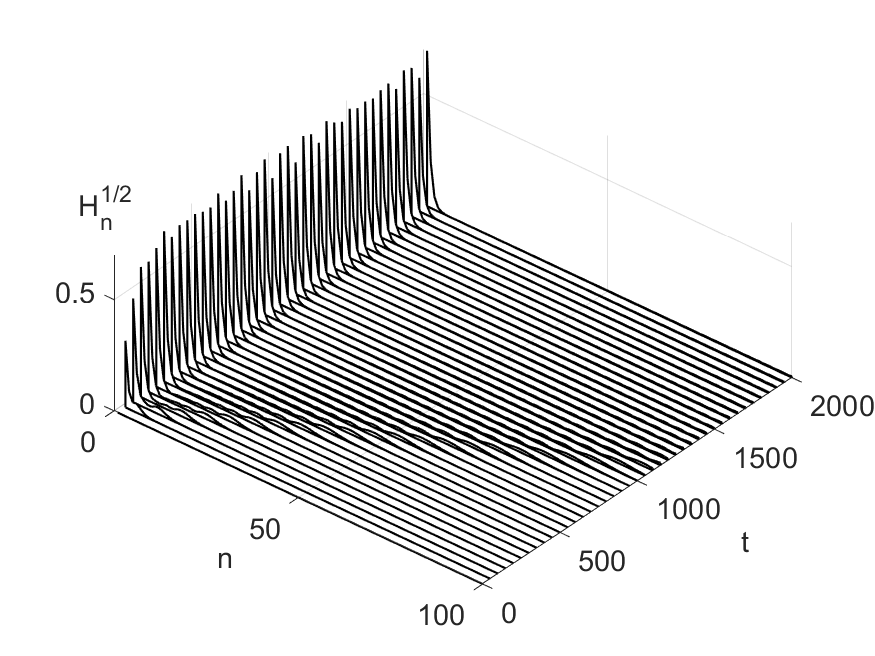}\label{subfig:supra_dynamics1_c}}
		\subfloat[$C=0.1$, $f=2.42$.]{\includegraphics[width=0.48\linewidth]{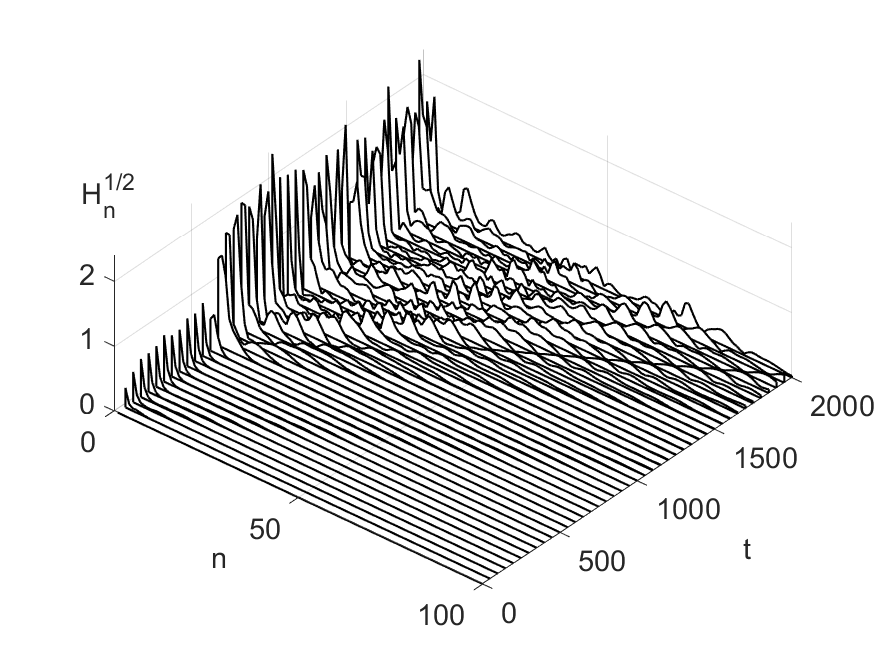}\label{subfig:supra_dynamics1_d}}
		\caption{
			Spatiotemporal evolution of $\sqrt{H_n}$ for the discrete sine--Gordon model \eqref{gov} with $p=1$ and $\omega=0.9$.  
			Panels (a,c): subthreshold forcing yields an evanescent response confined near the driven boundary.  
			Panels (b,d): slightly above threshold, the boundary excitation generates a propagating wavefront, indicating supratransmission.  
			For weak coupling ($C=0.1$), the transmitted energy is significantly reduced due to discreteness-induced trapping.}
		\label{fig:supra_dynamics}
	\end{figure}
	
	\begin{figure}[tbhp!]
		\centering
		\subfloat[$C=5$, $f=0.97$.]{\includegraphics[width=0.44\linewidth]{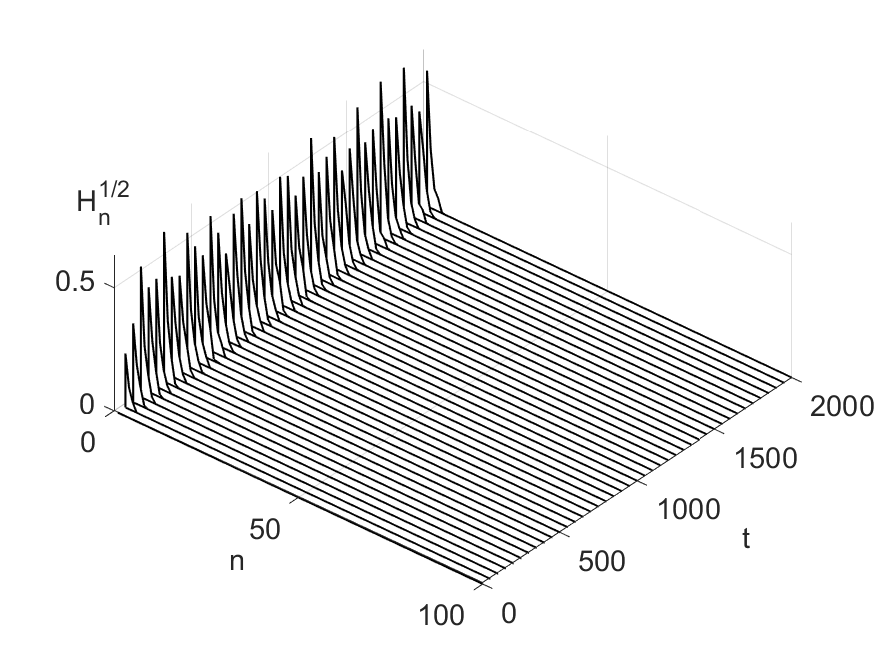}\label{subfig:supra_dynamics2_a}}
		\subfloat[$C=5$, $f=0.98$.]{\includegraphics[width=0.44\linewidth]{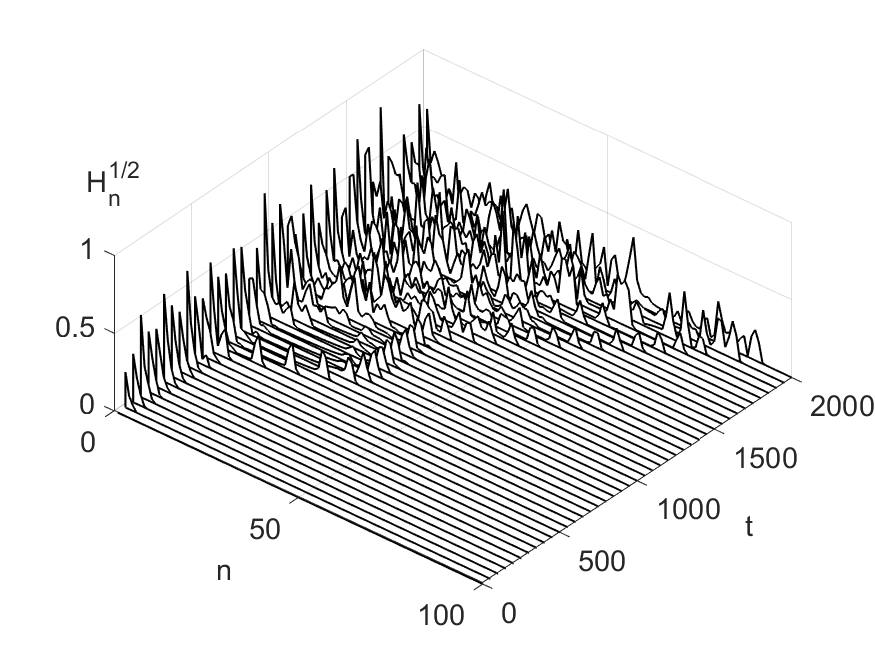}\label{subfig:supra_dynamics2_b}}\\
		\subfloat[$C=5$, $f=1.44$.]{\includegraphics[width=0.44\linewidth]{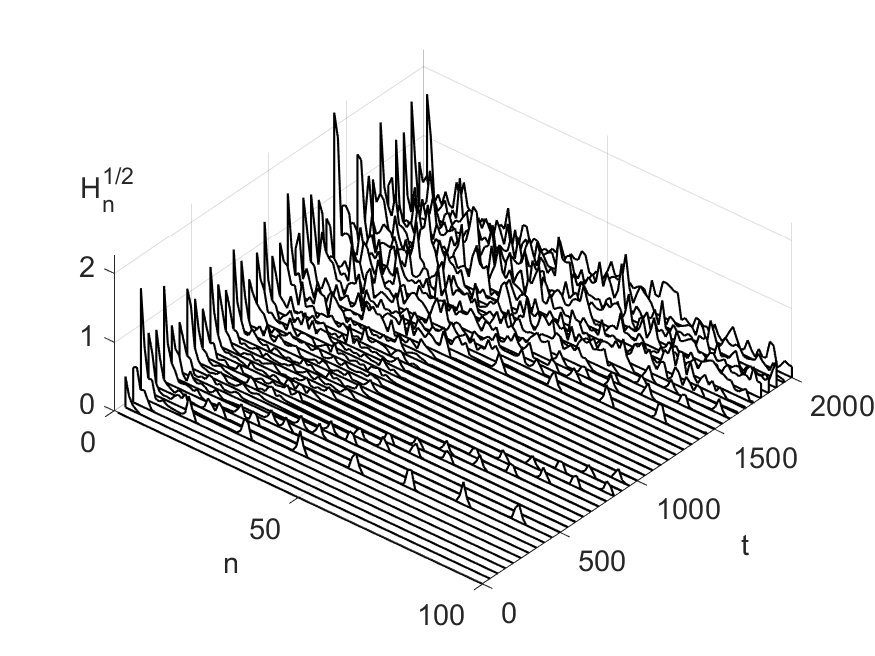}\label{subfig:supra_dynamics2_c}}
		\subfloat[$C=5$, $f=1.45$.]{\includegraphics[width=0.44\linewidth]{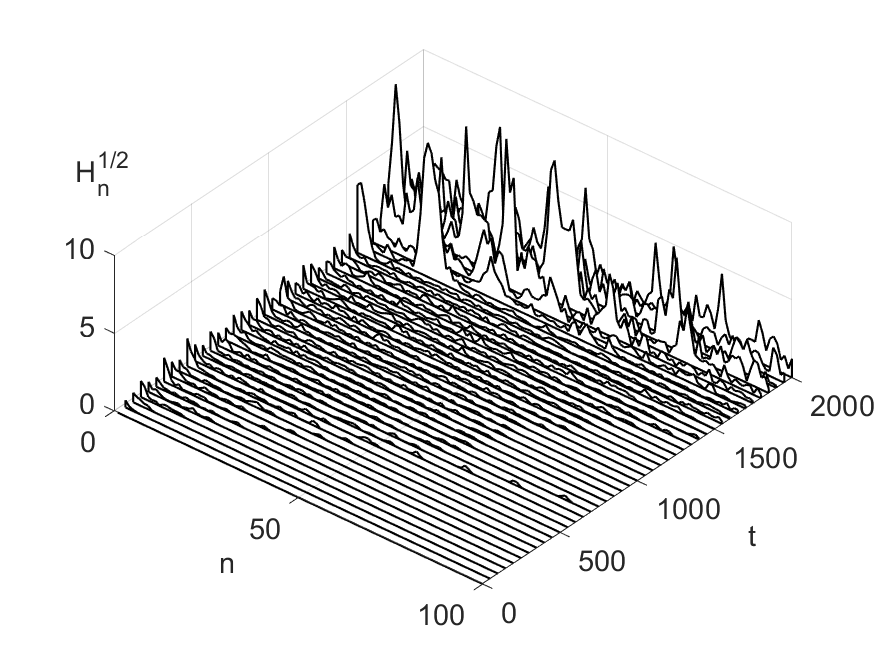}\label{subfig:supra_dynamics2_d}}\\
		\subfloat[$C=0.1$, $f=2.22$.]{\includegraphics[width=0.44\linewidth]{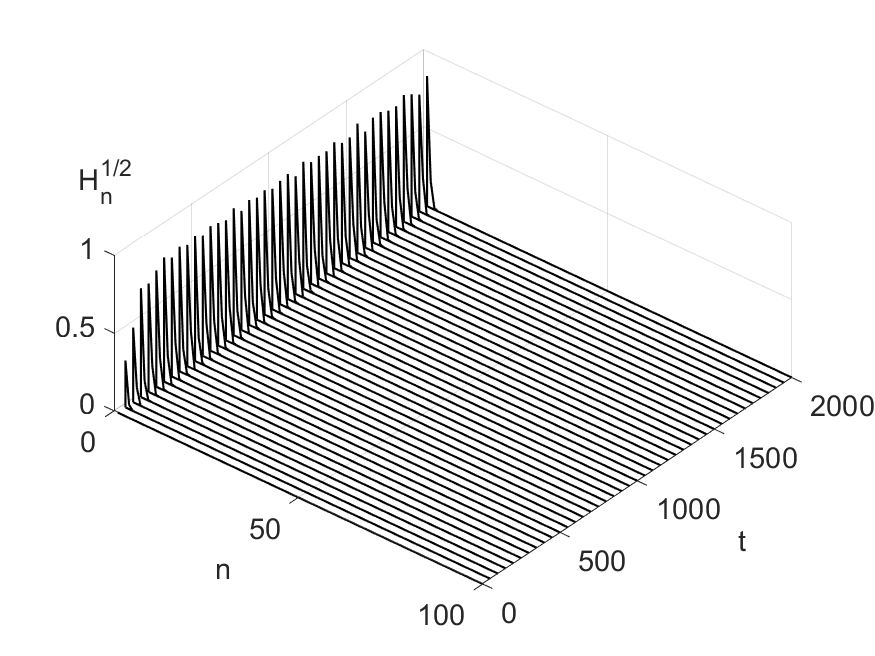}\label{subfig:supra_dynamics2_e}}
		\subfloat[$C=0.1$, $f=2.23$.]{\includegraphics[width=0.44\linewidth]{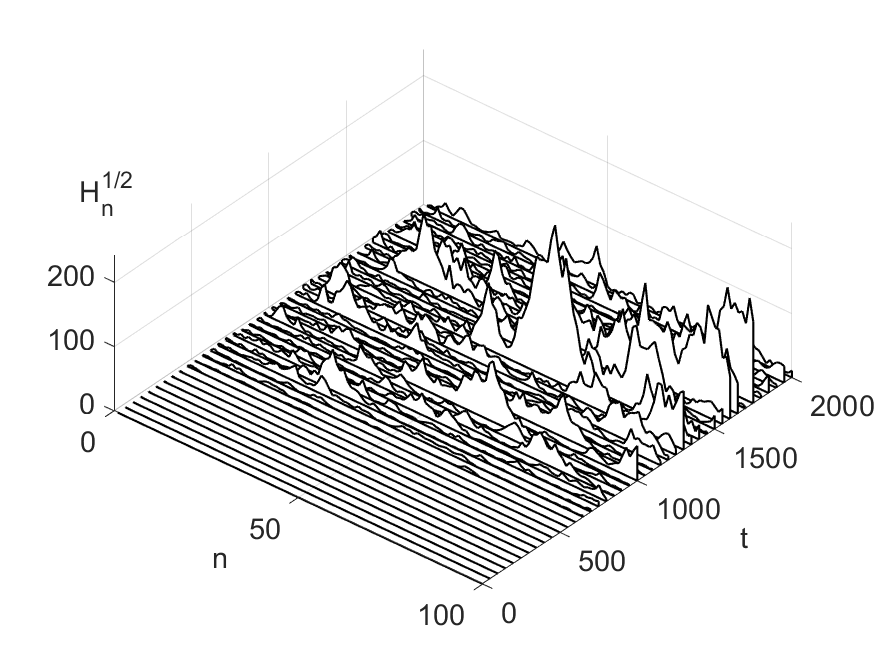}\label{subfig:supra_dynamics2_f}}
		\caption{
			Same setting as Fig.~\ref{fig:supra_dynamics}, but for nonlinear coupling with $p=3$.  
			Below threshold (a,c,e), the response remains evanescent.  
			Panels (b,c) show intermittent leakage due to weak instabilities of the evanescent state, but not sustained transmission.  
			Above threshold (d,f), pronounced energy propagation occurs, with stronger and more irregular wavefronts than in the linear case ($p=1$).}
		\label{fig:supra_dynamics2}
	\end{figure}
	
	We simulate the driven system \eqref{gov}--\eqref{drive} using a fourth-order Runge--Kutta scheme with time step $h=10^{-2}$.  {Convergence tests with smaller time steps (performed for representative parameter sets) confirmed that this choice provides sufficient accuracy for the reported results.}
	The lattice consists of $N=300$ sites.  
	To suppress reflections from the right boundary, a damping profile increasing smoothly over the last $50$ sites is applied.  
	The boundary forcing is ramped according to
	\begin{equation}
		F(t) = \bigl(1 - e^{-t/\tau}\bigr) f,
		\label{drive2}
	\end{equation}
	with $\tau=50$ to avoid initial transients.  
	Figures~\ref{fig:supra_dynamics} and \ref{fig:supra_dynamics2} display the square root of the local energy density $\sqrt{H_n}$ as a function of $n$ and $t$ for various combinations of $C$, $p$, and $f$ at fixed $\omega=0.9$. {The simulations are shown up to $T=2000$, which is sufficient to capture the long-time behavior of the system. Additional simulations over longer time intervals confirmed that no qualitative changes or delayed transmission occurred beyond this time window.}
	
	Figure~\ref{fig:supra_dynamics} shows the case $p=1$.  
	For both $C=5$ and $C=0.1$, the system supports evanescent waves when $f<f_{\mathrm{cr}}$, reproducing the classical behavior described in \cite{geniet2002energy}.  
	Panels~\ref{subfig:supra_dynamics1_a} and \ref{subfig:supra_dynamics1_c} show rapid spatial decay of the boundary excitation.  
	A slight increase of $f$ above threshold leads to propagation, as seen in Fig.~\ref{subfig:supra_dynamics1_b}.  
	For weak coupling ($C=0.1$), the transmitted energy is small and portions of the input remain trapped near the boundary, as in Fig.~\ref{subfig:supra_dynamics1_d}.  
	This trapping persists even for significantly larger amplitudes.
	
\begin{figure}[tbhp!]
		\centering
		\subfloat[$C=0.1$]{\includegraphics[width=0.48\linewidth]{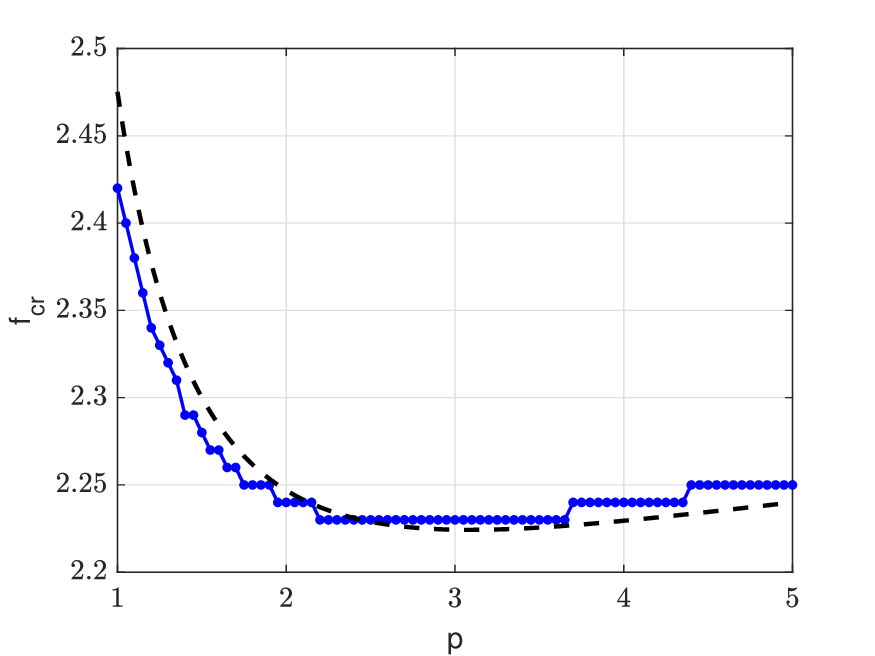}\label{subfig:p_vs_Fcrit_c01}}%
		\hfill
		\subfloat[$C=1$]{\includegraphics[width=0.48\linewidth]{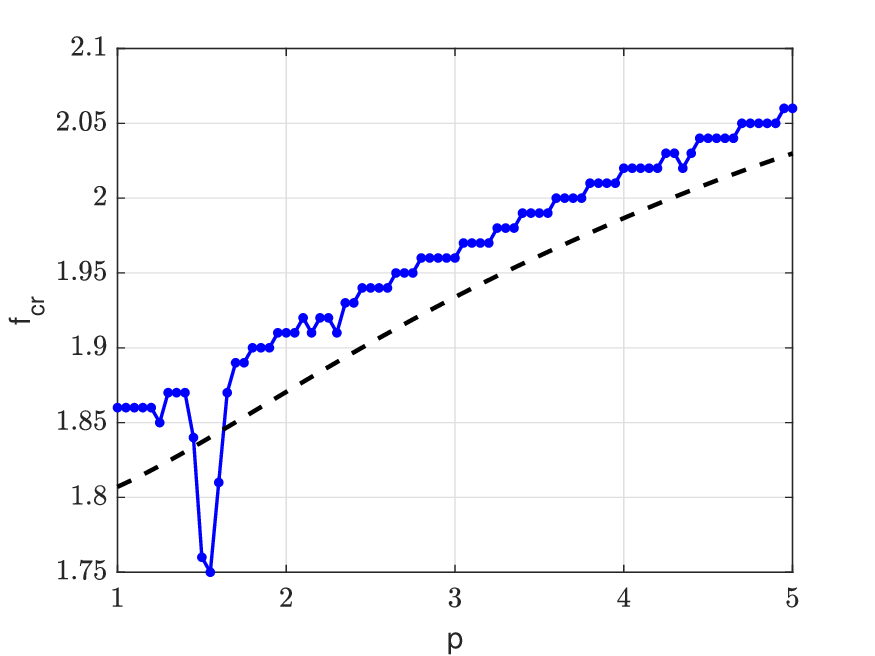}\label{subfig:p_vs_Fcrit_c1}}\\
		\subfloat[$C=5$]{\includegraphics[width=0.48\linewidth]{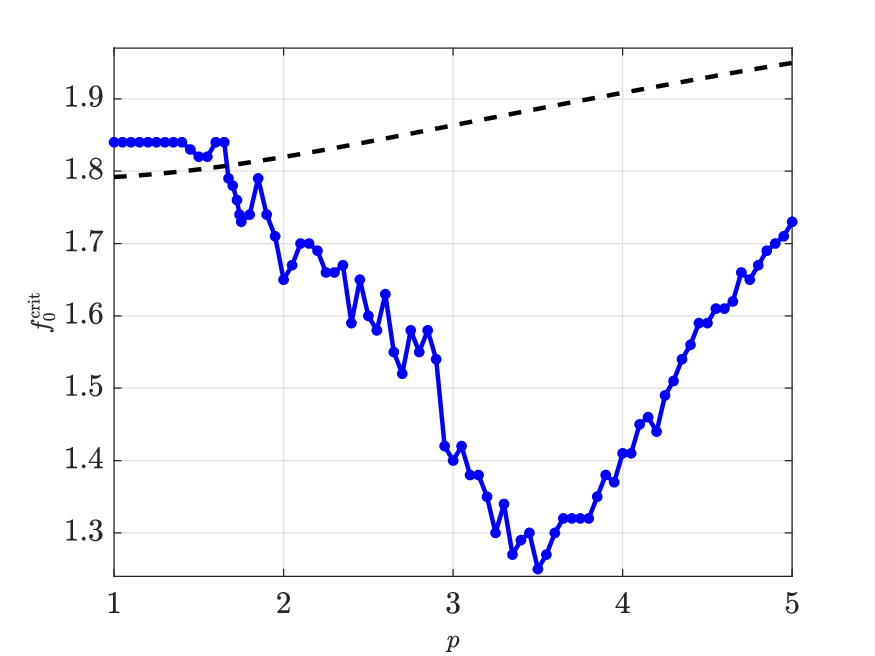}\label{subfig:p_vs_Fcrit_c5}}
		\caption{
			Critical driving amplitude $f_{\mathrm{cr}}$ versus coupling exponent $p$ for $\omega=0.9$ and three values of $C$.  
			Solid curves: thresholds obtained from the full model \eqref{gov}.  
			Dashed curves: predictions of the DpS approximation \eqref{newgov}.  
			For small and intermediate coupling, the DpS model reproduces the numerical trends; for strong coupling ($C=5$), it fails to capture the non-monotone dependence in the strong-coupling regime of the full model.}
		\label{fig:Fcrit_p}
	\end{figure}

	Figure~\ref{fig:supra_dynamics2} presents the corresponding results for $p=3$.  
	Below threshold (panels~\ref{subfig:supra_dynamics2_a} and \ref{subfig:supra_dynamics2_e}), the dynamics remain evanescent.  
	In panels~\ref{subfig:supra_dynamics2_b} and \ref{subfig:supra_dynamics2_c}, small intermittent bursts appear, but these do not lead to large energy propagation; rather, they are consistent with transient destabilization of the localized evanescent state. This behavior is reminiscent of the weak-transmission regime observed beyond the first threshold $F_{c1}$ in the nonlinear response manifold analysis of Maniadis \emph{et al.}~\cite{maniadis2006energy}, where a fold bifurcation destroys the stable localized periodic response and gives rise to weak radiative transmission. In the present setting, however, the mechanism differs substantially: the system is strongly coupled and the transition is mediated by nonlinear intersite coupling rather than by weakly coupled boundary-driven Klein--Gordon dynamics. Moreover, unlike the threshold $F_{c1}$ in \cite{maniadis2006energy}, the intermittent bursts observed here do not appear to be associated with a distinct fold structure in the continuation diagram, as we will discuss below (see Fig.~\ref{subfig:p_vs_Fcrit_c5}).
	Substantial energy transmission emerges only when $f>f_{\mathrm{cr}}$, as shown in Figs.~\ref{subfig:supra_dynamics2_d} and \ref{subfig:supra_dynamics2_f}. This transition is qualitatively similar to the second threshold $F_{c2}$ identified in \cite{maniadis2006energy}, which marks the onset of strong nonlinear energy transport. Compared to the linear-coupling case ($p=1$), the wavefronts for $p=3$ are more irregular and exhibit stronger amplitude fluctuations. For weak coupling, the transmitted energy is substantially larger than in Fig.~\ref{subfig:supra_dynamics1_d}, suggesting that nonlinear coupling enhances the efficiency of energy transport above threshold.
	
	Figure~\ref{fig:Fcrit_p} summarizes the dependence of the threshold amplitude $f_{\mathrm{cr}}$ on the exponent $p$ for weak, intermediate, and strong coupling.  {These trends indicate that the interplay between $C$ and $p$ is nontrivial, and that the transmission threshold is sensitive to both the interaction strength and the nonlinear coupling law. While the overall trends appear qualitatively similar across coupling regimes, the underlying mechanisms differ, as discussed below.}
	
	For $C=0.1$, Fig.~\ref{subfig:p_vs_Fcrit_c01}, the threshold decreases with increasing $p$, reaches a shallow minimum near $p\approx3$, and then rises.  
	At $C=1$ (Fig.~\ref{subfig:p_vs_Fcrit_c1}), the dependence becomes nearly linear, with small irregularities for $1<p<2$.  
	For $C=5$ (Fig.~\ref{subfig:p_vs_Fcrit_c5}), the behavior is strongly non-monotonic: $f_{\mathrm{cr}}$ decreases markedly up to $p\approx3.5$ and then increases.  
	These trends indicate that the interplay between $C$ and $p$ is nontrivial, and that the transmission threshold is sensitive to both the interaction strength and the nonlinear coupling law.
	
	The dashed curves in Fig.~\ref{fig:Fcrit_p} show the threshold predicted by the DpS reduction~\eqref{newgov} with boundary forcing~\eqref{drive3}.  
	For weak coupling ($C=0.1$), the DpS approximation agrees closely with the full system and reproduces the non-monotone dependence on $p$.  
	At $C=1$, the approximation remains accurate and captures the almost linear growth of $f_{\mathrm{cr}}$.  
	For strong coupling ($C=5$), however, the DpS model predicts a monotone increase of $f_{\mathrm{cr}}$, in contrast to the full dynamics.  
	{This discrepancy indicates a limitation of the envelope approximation in the strong-coupling regime. In particular, while the DpS model accurately predicts the critical amplitude in the weak-coupling regime, it fails to capture $f_{\mathrm{cr}}$ at strong coupling, suggesting a transition from a fold-induced to an instability-driven mechanism.}
	
	\begin{figure}[tbhp!]
		\centering
		\includegraphics[width=0.65\linewidth]{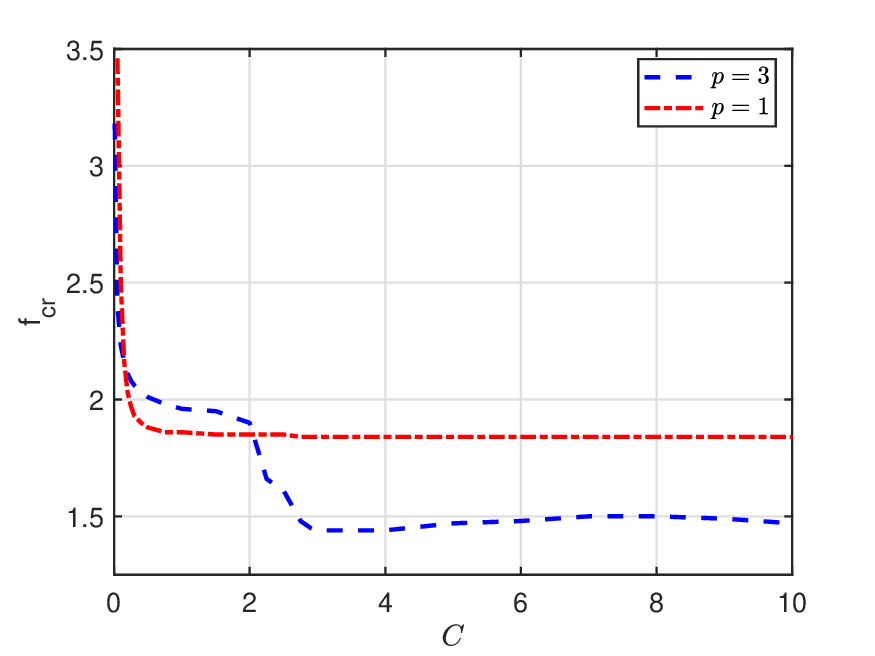}
		\caption{
			Critical driving amplitude as a function of coupling strength ${C}$ for $p=1$ and $p=3$.  
			In both cases, $f_{\mathrm{cr}}$ saturates as ${C}\to\infty$. For $p=1$, the limiting threshold is in agreement with the strong-coupling prediction of the DpS model \eqref{strcamp}. For $p=3$, it is slightly lower, reflecting the destabilizing effect of nonlinear coupling.}
		\label{Fig5}
	\end{figure}
	
	To examine the behavior of the supratransmission threshold in the strong-coupling regime, Fig.~\ref{Fig5} shows the dependence of the numerically determined critical amplitude $f_{\mathrm{cr}}$ on $C$ for the linear case ($p=1$) and for a strongly nonlinear coupling law ($p=3$).  
	For $p=1$, the threshold tends toward a constant value $f_{\mathrm{cr}}\approx 1.84$, in close agreement with the DpS prediction discussed later (Eq.~\eqref{strcamp}), which yields $f_{\mathrm{cr}}\approx 1.79$.  
	For $p=3$, a similar saturation occurs but at a lower level, consistent with the result in Fig.~\ref{subfig:p_vs_Fcrit_c5}, where supratransmission in the full model may be triggered by instabilities of the evanescent branch rather than a fold mechanism.

	\section{DpS as a small-amplitude approximation}\label{sec3}
	
	To obtain an analytical description of the numerical results in Sec.~\ref{sec2}, we derive the DpS equation \cite{james2011nonlinear}. We focus on small-amplitude, time-periodic solutions of period $2\pi$ and introduce a small parameter $\epsilon>0$ to capture slow temporal modulation.
	
	We write
	\begin{equation}
		y_n = Y_n^\epsilon(\tau,t), \qquad \tau = \epsilon^{2} t,
		\label{sol1}
	\end{equation}
	and adopt the ansatz
	\begin{equation}
		Y_n^\epsilon(\tau,t)
		= \epsilon\!\left(A_n^\epsilon(\tau)e^{i t}
		+ \overline{A}_n^\epsilon(\tau)e^{-i t}\right)
		+ \epsilon^{3} R_n^\epsilon(\tau,t),
		\label{sol2}
	\end{equation}
	where $A_n^\epsilon(\tau)$ is a slowly varying complex amplitude and $R_n^\epsilon$ is a real-valued remainder satisfying
	\[
	\int_0^{2\pi} R_n^\epsilon(\tau,t)\, e^{\pm i t}\, dt = 0.
	\]
	
	Substitution of \eqref{sol2} into \eqref{gov} and expansion of the amplitudes,
	\[
	A_n^\epsilon = A_n + o(\epsilon), \qquad
	R_n^\epsilon = R_n + o(\epsilon),
	\]
	gives an equation containing harmonics $e^{\pm i t}$ and higher-order contributions.  
	After discarding terms of order $o(\epsilon^{2})$, multiplying by $e^{-i t}$, and averaging over one period, only two types of nonlinear contributions remain: those arising from the coupling term and those from the onsite nonlinearity.
	
	Let $A_{n+1}-A_n = r e^{i\theta}$.  
	The averaged coupling contribution involves integrals of the form
	\[
	\int_0^{2\pi}
	\operatorname{sign}\big(re^{i(t+\theta)} + \text{c.c.}\big)
	\left| re^{i(t+\theta)} + \text{c.c.} \right|^{p} e^{-i t}\, dt.
	\]
	Only the real part survives, yielding the Wallis-type integral
	\[
	\int_0^{2\pi} |\cos(t+\theta)|^{p+1} dt
	= \frac{2\sqrt{\pi}\,\Gamma(\tfrac{p}{2}+1)}{\Gamma(\tfrac{p+1}{2}+1)}.
	% =: K_p.
	\]
	The imaginary part vanishes by symmetry. For the cubic onsite contribution, $(e^{i t}A_n + \text{c.c.})^3$, only the projection onto $e^{i t}$ survives the averaging, giving $6\pi\, r^{3} e^{i\theta}$.
	
	The above averaged terms lead to the discrete $p$-Schr\"odinger equation
	\begin{equation}
		i\,\partial_\tau A_n
		= \tilde{C}
		\Delta_p A_n
		+ \tilde{\gamma}|A_n|^2 A_n,
		\label{newgov}
	\end{equation}
	where 
	\begin{align*}
		\Delta_{p}f_n&:=(f_{n+1}-f_n)|f_{n+1}-f_n|^{p-1}\\
		&\quad+(f_{n-1}-f_n)|f_{n-1}-f_n|^{p-1}
	\end{align*}
	and the effective coefficients are
	\begin{align}
		&\tilde{C}
		= \epsilon^{p-3}\,
		\frac{C\,c_p\,2^{\,p-2}}{\pi},\quad
		c_p
		= \frac{2\sqrt{\pi}\, p\, \Gamma(\tfrac{p}{2})}
		{(p+1)\Gamma(\tfrac{p+1}{2})},\quad
		\tilde{\gamma} = \frac{3}{2}\gamma.\nonumber
		\label{coeff}
	\end{align}
	Equation (\ref{newgov}) serves as the reduced model for the subsequent analytical investigation of the transmission threshold. 
	
	\begin{figure}[tbhp!]
		\centering
		\includegraphics[width=0.65\linewidth]{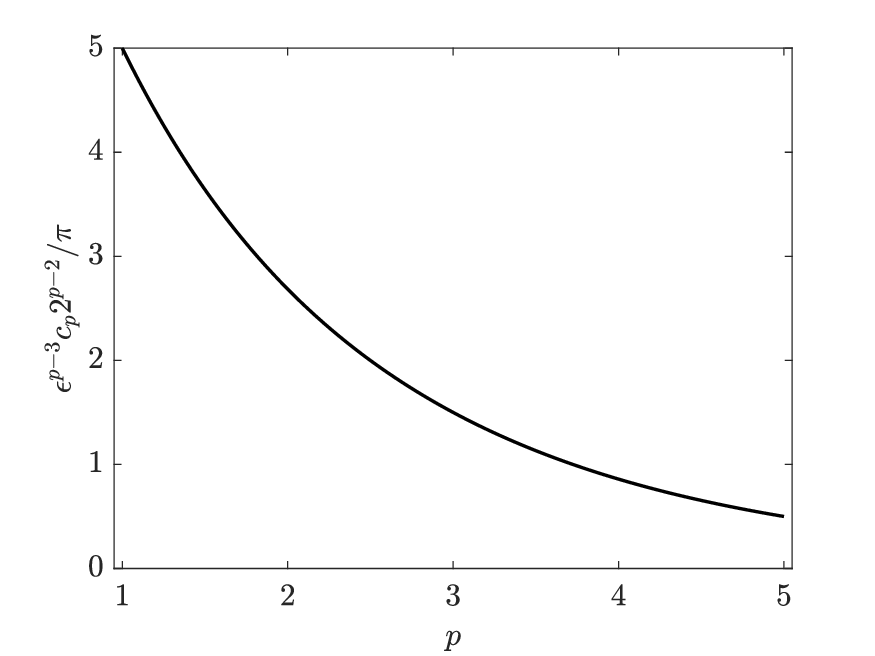}
		\caption{
			Dependence of the effective coupling prefactor $\tilde{C}/C$ on the exponent $p$ for $\omega=0.9$ and $\Omega=1$.  
			The monotone decrease indicates weaker envelope coupling in the DpS reduction as $p$ increases.}
		\label{fig:p_vs_cp}
	\end{figure}
	
	Figure~\ref{fig:p_vs_cp} shows the ratio $\tilde{C}/C$ as a function of $p$.  
	The monotone decrease highlights that the envelope dynamics become increasingly weakly coupled for larger coupling exponents.
	
	The boundary forcing must be expressed at the amplitude level.  
	From \eqref{sol2}--\eqref{drive2}, the leading-order contribution gives
	\begin{equation}
		A_0 = \frac{F}{2\epsilon}\, e^{-i\Omega\tau},
		\label{drive3}
	\end{equation}
	where
	\begin{equation}
		\epsilon = \sqrt{\frac{1-\omega}{\Omega}}.
		\label{eps}
	\end{equation}
	Without loss of generality, we set $\Omega = 1$ in the following.

	\section{Evanescent waves and their stability}\label{sec4}
	
	We first recall the basic plane-wave properties of the DpS equation~\eqref{newgov} on the infinite lattice $n\in\mathbb{Z}$.  
	Looking for Bloch-type solutions of the form
	\begin{equation}
		A_n(\tau) = \tilde{a}\,e^{i(-\Omega \tau + q n + \phi)}, 
		\qquad q\in[-\pi,\pi],
		\label{twsol}
	\end{equation}
	with constant complex amplitude $\tilde{a}$ and wavenumber $q$, we obtain the nonlinear dispersion relation
	\begin{equation}
		\Omega 
		= -\,\tilde{C}\,2^{p+1}\,|\tilde{a}|^{\,p-1}
		\bigl|\sin\tfrac{q}{2}\bigr|^{\,p+1}
		+ \tilde{\gamma}\,|\tilde{a}|^{2}.
		\label{disper2}
	\end{equation}
	In the linear case $p=1$ and for small amplitudes $|\tilde{a}|\to 0$, the onsite term becomes negligible and we recover
	\[
	\Omega = -4\tilde{C}\,\sin^2\!\tfrac{q}{2}.
	\]
	Since $\sin^2(q/2)\in[0,1]$, the corresponding frequency band for real wavenumbers is
	\[
	\Omega \in [-4\tilde{C},\,0].
	\]
	Outside this interval the wavenumber $q$ is complex and the corresponding modes are spatially decaying (evanescent).
	
	\subsection{Standing evanescent waves}
	
	We now consider the boundary-driven DpS equation on a semi-infinite lattice $n\in\mathbb{N}$ and seek standing waves oscillating at the driving frequency.  
	Motivated by the boundary condition~\eqref{drive3}, we use the ansatz
	\begin{equation}
		A_n(\tau) = a_n\,e^{i(-\Omega\tau + \phi)},
	\end{equation}
	where $a_n\in\mathbb{R}$ is time-independent and we impose
	\begin{equation}
		a_n \to 0 \quad \text{as}\quad n\to\infty,
	\end{equation}
	so that the resulting solution is evanescent.
	
	Substituting into \eqref{newgov}, we obtain the stationary boundary-value problem
	\begin{equation}
		\Omega a_n
		= \tilde{C}\Delta_p a_n
		+ \tilde{\gamma}a_n^{3},
		\label{govtra}
	\end{equation}
	with the boundary condition
	\begin{equation}
		a_0 = \frac{F}{2\epsilon}.
		\label{drive1}
	\end{equation}
	For given $(\Omega,F)$ and parameters $(p,\tilde C,\tilde\gamma)$, \eqref{govtra}--\eqref{drive1} admits families of evanescent profiles $a_n$ that will play a central role in the onset of supratransmission.
	
	\subsection{Linear stability analysis}
	
	We next examine the linear stability of a given standing-wave solution $\{a_n\}$.  
	To this end, we perturb $A_n(\tau)$ as
	\begin{equation}
		A_n(\tau)
		= \bigl[a_n + \xi\,(u_n(\tau) + i v_n(\tau))\bigr]
		e^{i(-\Omega\tau + \phi)},
	\end{equation}
	where $u_n(\tau)$ and $v_n(\tau)$ are real-valued perturbation components, and small parameter $0<\xi\ll 1$.  
	Substituting into \eqref{newgov}, expanding in $\xi$, and collecting terms yields, at order $\mathcal{O}(\xi)$, a linear system for $(u_n,v_n)$.
	
	For the coupling term we expand
	\[
	(A_{n+1}-A_n)\,|A_{n+1}-A_n|^{p-1}
	\]
	around the stationary profile.  
	Writing $d_n := a_{n+1}-a_n$, a Taylor expansion in the small increment (using the generalized binomial theorem)
	\[
	\xi\big[(u_{n+1}-u_n) + i(v_{n+1}-v_n)\big]
	\]
	gives, up to $\mathcal{O}(\xi)$,
	\begin{align}
		\nonumber&(A_{n+1}-A_n)\,|A_{n+1}-A_n|^{p-1}\\
		\nonumber&= \Big[
		d_n^{\,p}
		+ \xi\,p\,d_n^{\,p-1}(u_{n+1}-u_n)\Big]e^{i(-\Omega\tau+\phi)}\\
		&\quad + \Big[i\xi\,d_n^{\,p-1}(v_{n+1}-v_n)
		+ \mathcal{O}(\xi^2)
		\Big] e^{i(-\Omega\tau+\phi)}.
		\label{eig2}
	\end{align}
	An analogous expression holds for $(A_{n-1}-A_n)|A_{n-1}-A_n|^{p-1}$ with $d_n$ replaced by $a_n-a_{n-1}$.
	
	For the onsite term we similarly obtain
	\begin{align}
		\nonumber A_n|A_n|^2&= \bigl(a_n + \xi(u_n+i v_n)\bigr)
		\bigl(a_n^2 + \mathcal{O}(\xi)\bigr)
		e^{i(-\Omega\tau+\phi)} \\
		&= \Big[a_n^3 + \xi\bigl(3a_n^2 u_n + i a_n^2 v_n\bigr)
		+ \mathcal{O}(\xi^2)\Big] e^{i(-\Omega\tau+\phi)}.
		\label{eig3}
	\end{align}
	
	Collecting $\mathcal{O}(\xi)$ terms and separating real and imaginary parts leads to the evolution system
	\begin{equation}
		\frac{d}{d\tau}
		\begin{bmatrix}
			u_n \\[2pt] v_n
		\end{bmatrix}
		=
		\underbrace{
			\begin{bmatrix}
				0      & L_{+} \\
				-L_{-} & 0
		\end{bmatrix}}_{\displaystyle \mathcal{L}}
		\begin{bmatrix}
			u_n \\[2pt] v_n
		\end{bmatrix},
		\label{linop}
	\end{equation}
	where the self-adjoint difference operators $L_{+}$ and $L_{-}$ acting on sequences $\{u_n\}$ and $\{v_n\}$ are defined as follows
	\[
	(\Delta^+ w)_n = w_{n+1} - w_n,
	\qquad
	(\Delta^- w)_n = w_n - w_{n-1},
	\]
	and the weights
	\[
	w_n^+ = |a_{n+1}-a_n|^{p-1}, \qquad
	w_n^- = |a_{n}-a_{n-1}|^{p-1}.
	\]
	Then the action of $L_{+}$ and $L_{-}$ can be written as
	\begin{align}
		\nonumber (L_{+} v)_n
		&= \tilde{C}\Big[
		(\Delta^+ (w^+ \Delta^+ v))_n
		+ (\Delta^- (w^- \Delta^- v))_n
		\Big]\\
		&\quad + (\tilde{\gamma} a_n^2 - \Omega)\,v_n,
		\label{Lplus-op} \\
		\nonumber (L_{-} u)_n
		&= p\,\tilde{C}\Big[
		(\Delta^+ (w^+ \Delta^+ u))_n
		+ (\Delta^- (w^- \Delta^- u))_n
		\Big]\\
		&\quad+ (3\tilde{\gamma} a_n^2 - \Omega)\,u_n.
		\label{Lminus-op}
	\end{align}
	
	The spectral properties of $\mathcal{L}$ determine the linear stability of the standing wave.  
	If at least one eigenvalue $\lambda$ of $\mathcal{L}$ satisfies $\text{Re}(\lambda)>0$, the corresponding perturbation grows exponentially and the standing wave is linearly unstable.  
	If all eigenvalues satisfy $\text{Re}(\lambda)=0$, the solution is linearly stable.  
	Complex eigenvalues with nonzero real part correspond to oscillatory instabilities, whereas purely real positive eigenvalues are associated with monotone (non-oscillatory) exponential growth.
	
	\section{Numerical results of the DpS}\label{sec5}
	
	\begin{figure}[tbhp!]
		\centering
		\subfloat[$C=0.1$]{\includegraphics[width=0.8\linewidth]{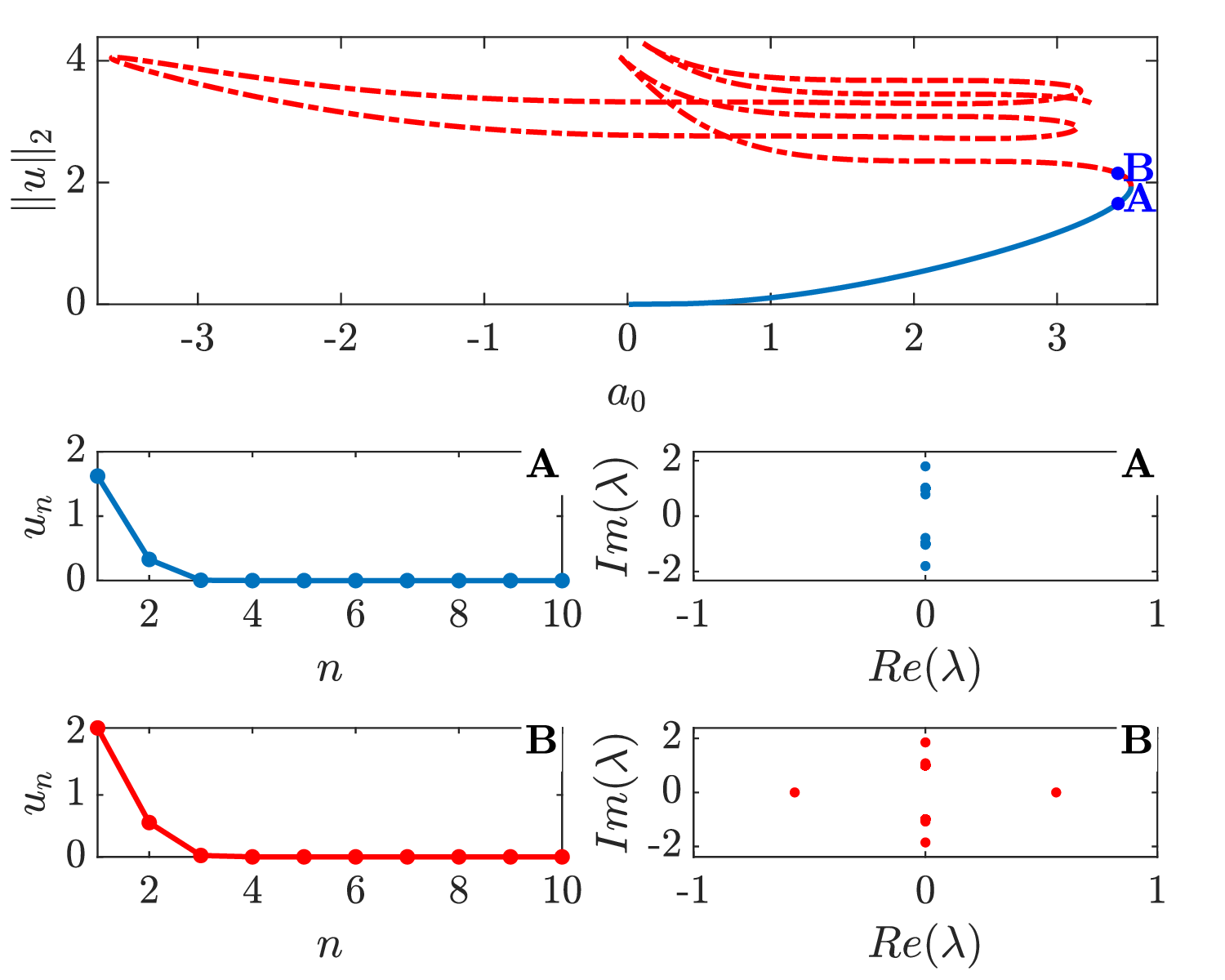}}\\
		\subfloat[$C=5$]{\includegraphics[width=0.8\linewidth]{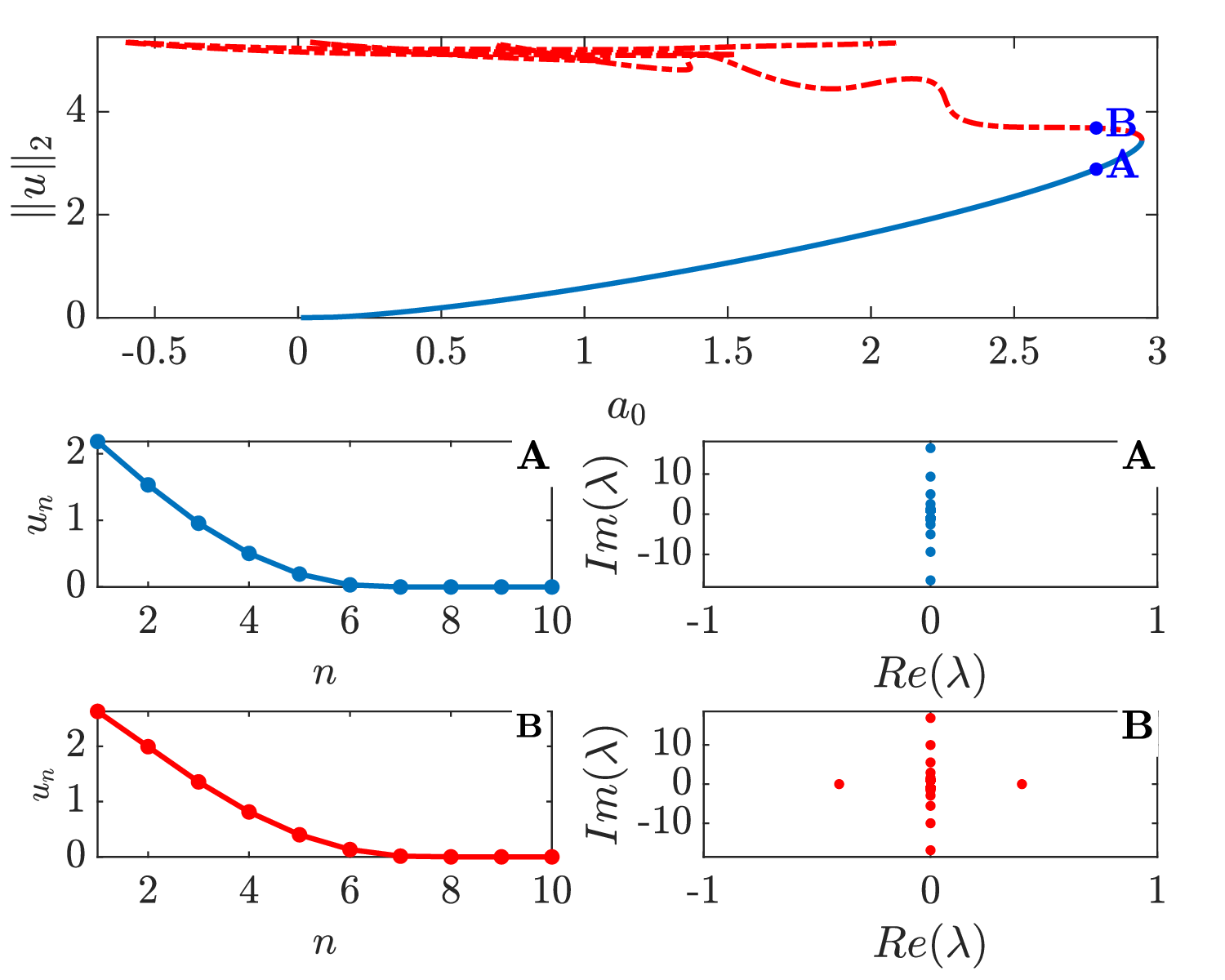}}
		\caption{
			Bifurcation structure of standing-wave solutions of the DpS equation \eqref{govtra} for $p=3$, $\Omega=1$, $\omega=0.9$, and $\tilde{\gamma}=1/4$.  
			Top panels: $\ell_2$-norm of the stationary solution versus driving amplitude $a_0$, with stable (solid) and unstable (dashed) branches.  
			A fold (saddle-node) bifurcation marks the supratransmission threshold.  
			Bottom panels: representative profiles and spectra at points A (stable, subthreshold) and B (unstable, above threshold).  
			Stronger coupling shifts the fold toward smaller $a_0$.}
		\label{Fig3}
	\end{figure}
	
	We compute localized stationary solutions of \eqref{govtra}--\eqref{drive1} using \texttt{fsolve} in \textsc{MATLAB}, treating the driving amplitude $a_0$ as the primary continuation parameter. Solution branches are followed through fold bifurcations via pseudo-arclength continuation. For each stationary state, linear stability is determined from the spectrum of the operator $\mathcal{L}$ defined in \eqref{linop}. The bifurcation structure is represented in terms of the $\ell_2$-norm
	\[
	\|u\|_2 = \Bigl(\sum_{n=1}^{N} |u_n|^2\Bigr)^{1/2}.
	\]
	The resulting continuation diagram is structurally related to the nonlinear response manifold framework introduced by Kopidakis and Aubry \cite{kopidakis2000discrete,kopidakis2000intraband} for localized harmonic forcing in discrete Klein--Gordon lattices. In particular, both settings exhibit folded solution branches with bifurcations. The nonlinear response manifold approach was later applied to nonlinear energy transmission and supratransmission phenomena in semi-infinite driven Klein--Gordon chains by Maniadis \emph{et al.} \cite{maniadis2006energy} and Johansson \emph{et al.} \cite{johansson2009transmission}, while related boundary-driven problems in the discrete nonlinear Schr\"odinger setting were subsequently investigated by Susanto \cite{susanto2008boundary}.
	
	Figure~\ref{Fig3} shows the bifurcation diagrams for two representative coupling strengths, $C=0.1$ and $C=5$, with $p=3$.  
	In both cases, the standing-wave branch exhibits a fold bifurcation: the lower branch is stable up to the fold, after which the solutions lose stability.  
	The location of the fold shifts toward smaller values of $a_0$ as $\tilde{C}$ increases, reflecting stronger effective coupling.
	
	Representative profiles and their spectra are shown below each bifurcation diagram.  
	At point A (just below threshold), the solution is spectrally stable.  
	At point B (just above threshold), a pair of real eigenvalues appears, indicating exponential instability.  
	This local loss of stability corresponds to the DpS prediction of supratransmission onset.
	
	The turning points of the DpS equation were computed for the same parameter sets used in the full simulations of the original model, shown in Fig.~\ref{fig:Fcrit_p}.  
	The corresponding threshold in the original system is obtained using the boundary relation \eqref{drive1}. The resulting critical amplitudes (dashed curves) agree very well with the full model for $C=0.1$ and $C=1$, confirming that the DpS reduction accurately captures the threshold in these regimes.  
	For $C=5$, however, the agreement breaks down: while the DpS model predicts a fold-induced threshold, the full system exhibits supratransmission at significantly smaller amplitudes.
	
	This indicates that, for strong coupling, the onset of transmission in the original equation \eqref{gov} does not appear to arise from a fold of the stationary branch. Instead, it suggests an instability-driven mechanism in which an evanescent state on the lower branch becomes unstable and initiates energy propagation once perturbations exceed a critical level (see \cite{susanto2023surge,kusdiantara2025band}).  
	The DpS model does not capture this effect because the lower branch remains spectrally stable in \eqref{newgov}.  
	A full continuation and stability analysis of evanescent waves in the original system would be required to characterize this strong-coupling regime, which is left for future work.
	
	\section{Analytical approximations of the critical amplitude}
	\label{sec6}
	
	To complement the numerical continuation results of Sec.~\ref{sec5}, we derive analytical approximations for the critical driving amplitude at which the stationary branch of the DpS equation folds.  
	The analysis proceeds in two regimes: a weak-coupling limit, where the solution is strongly localized and permits a perturbative expansion, and a strong-coupling limit, where a continuum-type scaling yields tractable asymptotics.  
	Both approaches provide insight into how the coupling strength and the exponent $p$ influence the supratransmission threshold.
	
	\subsection{Weak-coupling limit}
	
	In the regime $\tilde{C}\ll1$, solutions of \eqref{govtra} are sharply localized and the leading sites dominate the balance determining the fold.  
	The stationary equation may therefore be treated site-by-site via a perturbative expansion in powers of $\tilde{C}$.
	
	\subsubsection{Approximate solutions for small driving}
	
	We introduce the series
	\begin{equation}
		a_i = a_i^{(i)} \tilde{C}^{\alpha_i}
		+ a_i^{(i+1)}\tilde{C}^{2\alpha_i}
		+ a_i^{(i+2)}\tilde{C}^{3\alpha_i}
		+ \dotsi,
		\label{solpert}
	\end{equation}
	with exponents $\alpha_i$ selected to balance the nonlinear and coupling terms.  
	For $p=1$ (linear coupling), choosing $\alpha_i=i$ and inserting \eqref{solpert} into \eqref{govtra} yields the expansions up to $\mathcal{O}(\tilde{C}^3)$:
	\begin{subequations}
		\begin{align}
			a_1 &= \frac{a_0}{\Omega}\tilde{C}
			- \frac{2a_0}{\Omega^2}\tilde{C}^2
			+ \left(\frac{5a_0}{\Omega^3}
			+ \frac{\tilde{\gamma}a_0^3}{\Omega^4}\right)\tilde{C}^3 + \dots,\\
			a_2 &= \frac{a_0}{\Omega^2}\tilde{C}^2
			- \frac{4a_0}{\Omega^3}\tilde{C}^3 + \dots,\\
			a_3 &= \frac{a_0}{\Omega^3}\tilde{C}^3 + \dots.
		\end{align}
	\end{subequations}
	The nonlinear correction first appears at $\mathcal{O}(\tilde{C}^3)$, proportional to $\tilde{\gamma}$.
	
	For $p>1$, monotonic decay ($a_{n-1} \geq a_n$) allows us to write
	\[
	(a_n - a_{n\pm1})|a_n - a_{n\pm1}|^{p-1}
	= \pm |a_n - a_{n\pm1}|^p,
	\]
	removing sign ambiguities.  
	The perturbation series requires consistent powers of $\tilde{C}$, which is possible only when $p$ is rational.  
	If $p=m/q$ in lowest terms, we choose $\alpha_i = i/q$.
	
	As an example, take $p=\tfrac{3}{2}$, so $\alpha_i=i/2$.  
	Applying the expansion procedure yields
	\begin{align}
		\nonumber a_1 &= \frac{a_0^{3/2}}{\Omega}\tilde{C}
		- \frac{3a_0^2}{2\Omega^2}\tilde{C}^2
		- \frac{a_0^{9/4}}{\Omega^{5/2}}\tilde{C}^{5/2}\\
		&\quad+ \left(\frac{21a_0^{5/2}}{8\Omega^3}
		+ \tilde{\gamma}\frac{a_0^{9/2}}{\Omega^4}\right)\tilde{C}^3 + \dots,
		\label{sola1p}\\
		a_2 &= \frac{a_0^{9/4}}{\Omega^{5/2}}\tilde{C}^{5/2} + \dots.
		\label{sola2p}
	\end{align}
	The excitation of higher sites occurs at higher powers of $\tilde{C}$ than in the linear case, reflecting stronger spatial localization for $p>1$.  
	As $p$ increases, the front penetrates more slowly into the lattice.
	
	\subsubsection{Approximate critical amplitude}
	
	We now derive an estimate for the fold point.  
	Balancing the nonlinear term with the coupling contributions yields the scalings
	\begin{align*}
		&a_0=\tilde{C}^{-\frac{1}{2(p-1)}}\hat{a}_0,\quad
		a_1=\tilde{C}^{\frac{p-2}{2(p-1)}}\hat{a}_1,\\
		&a_2=\tilde{C}^{\frac{2p-3}{2(p-1)}}\hat{a}_2,\quad
		\tilde{\gamma}=\tilde{C}^{-\frac{p-2}{p-1}}\hat{\gamma}.
	\end{align*}
	Substitution into \eqref{govtra} yields
	\begin{equation}
		-\Omega \hat{a}_1
		- \tilde{C}^{p/2}(\hat{a}_1-\hat{a}_2\sqrt{\tilde{C}})^p
		+ (\hat{a}_0-\hat{a}_1\sqrt{\tilde{C}})^p
		+ \hat{\gamma} \hat{a}_1^3 = 0.
		\label{crita0p}
	\end{equation}
	Matching powers of $\tilde{C}$ gives the leading-order approximation
	\begin{equation}
		\hat{a}_0^{\mathrm{cr}}
		= \left(\frac{2\Omega\sqrt{3\hat{\gamma}\Omega}}{9\hat{\gamma}}\right)^{1/p}
		+ \sqrt{\frac{\Omega}{3\hat{\gamma}}}\,\tilde{C}+\dots,
		\label{eq:approx}
	\end{equation}
	which provides an explicit estimate for the critical amplitude. The leading term corresponds to the uncoupled limit, while the correction term quantifies the shift induced by weak coupling. We will compare this result with numerical approximation using arclength continuation in Sec.~\ref{sec5}.

	\subsection{Strong-coupling limit}
	
	\begin{figure}[tbhp]
		\centering
		\subfloat[]{\includegraphics[width=0.48\linewidth]{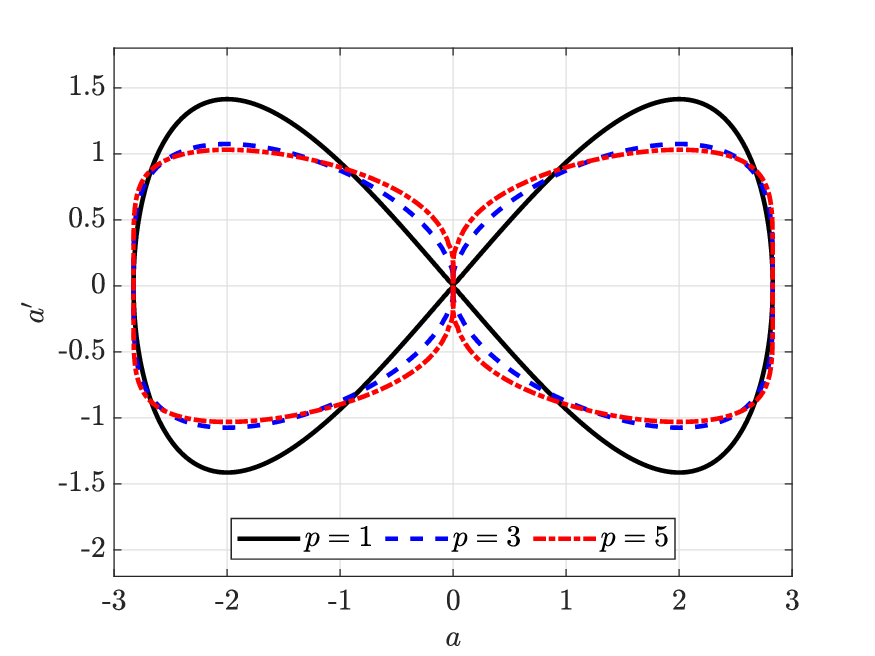}}
		\subfloat[]{\includegraphics[width=0.48\linewidth]{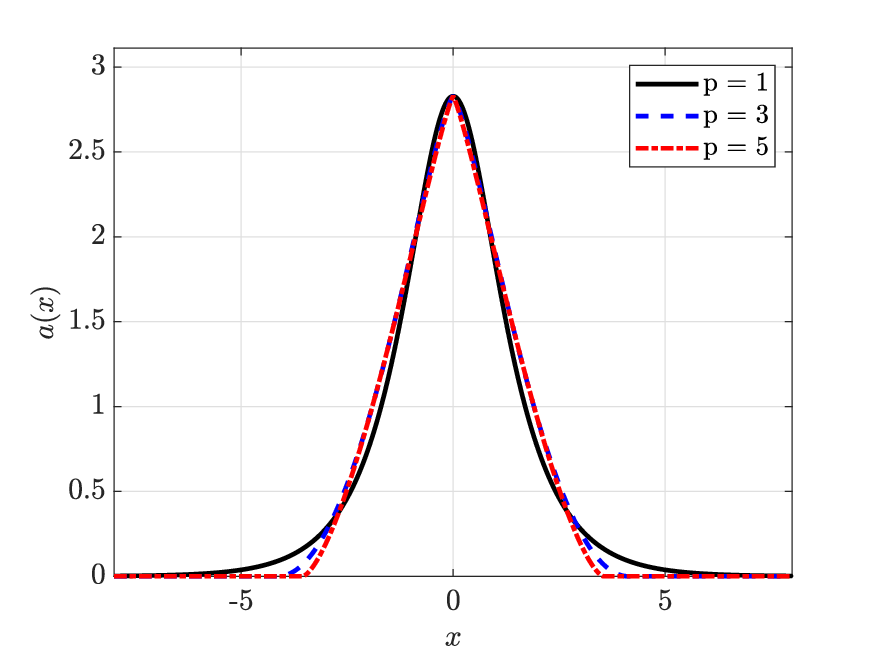}}
		\caption{
			(a) Phase-plane trajectories of the continuum strong-coupling model \eqref{traj} for several values of $p$ and parameters $\Omega=1$, $\tilde{\gamma}=1/4$.  
			(b) Corresponding localized profiles obtained from \eqref{sol_traj}.  
			The amplitude is independent of $p$, consistent with \eqref{strcamp}.}
		\label{fig:traj_n_Sol}
	\end{figure}
	
	We now consider the strong-coupling regime of the DpS equation, corresponding to a continuum limit in which many lattice sites are excited and the effective spacing is small.  
	Introduce a small parameter $\varepsilon>0$ and the continuum variable $x=\varepsilon n$.  
	Assuming a strong-coupling scaling $\tilde{C} = \varepsilon^{-(p+1)}$ and taking the formal limit $\varepsilon\to0$, Eq.~\eqref{newgov} becomes the quasilinear NLS-type equation
	\begin{equation}
		iA_{\tau} = \bigl(A_{x}|A_{x}|^{p-1}\bigr)_{x} + \tilde{\gamma} A|A|^2.
		\label{newgov4}
	\end{equation}
	Thus, strong coupling manifests in a nonlinear dispersive term, while the onsite nonlinearity remains unchanged.
	
	For later use, we also consider the limit $p\to1^{+}$ by setting $p-1=\xi$ with $\xi\to0^{+}$.  
	Expanding \eqref{newgov4} in $\xi$ gives
	\begin{equation}
		iA_{\tau} 
		= A_{xx}\Big(1 + \xi + \xi\ln|A_x|\Big) 
		+ \tilde{\gamma} A|A|^2,
		\label{newgov5}
	\end{equation}
	which recovers the standard cubic NLS dispersion as $\xi\to0$.
	
	We now seek standing waves of the form
	\[
	A(x,\tau) = a(x)e^{i(-\Omega\tau+\phi)},
	\]
	with real-valued $a(x)$.  
	Assuming a symmetric, positive profile with $a'(x)\le0$ for $x>0$, substitution into \eqref{newgov4} yields
	\begin{equation}
		\Omega a = p(-a')^{p-1} a'' + \tilde{\gamma} a^3.
	\end{equation}
	Multiplying by $a'$ and integrating once gives the first integral
	\begin{equation}
		\frac{1}{2}\Omega a^2 - \frac{p}{p+1}(-a')^{p+1} - \frac{1}{4}\tilde{\gamma} a^4 = 0.
		\label{traj}
	\end{equation}
	The corresponding phase-plane trajectories and localized profiles are shown in Fig.~\ref{fig:traj_n_Sol} for several values of $p$.  
	
	Following the standard argument for boundary-driven NLS-type systems (see, e.g., \cite{geniet2002energy}), the critical driving amplitude coincides with the peak amplitude of the localized wave.  
	Setting $a'=0$ in \eqref{traj} yields
	\begin{equation}
		a^{\mathrm{cr}} = \sqrt{\frac{2\Omega}{\tilde{\gamma}}},
		\label{strcamp}
	\end{equation}
	which is independent of $p$ within the continuum approximation and is valid for both \eqref{newgov4} and its near-linear limit \eqref{newgov5}.
	
	To obtain the soliton envelope, we return to \eqref{traj}.  
	For $p=1$, one recovers the usual $\sech$-type profile.  
	For $p>1$, we have
	\[
	-a' = \left[\frac{p+1}{p} a^2 \left(\frac{\Omega}{2} - \frac{\tilde{\gamma} a^2}{4}\right)\right]^{\frac{1}{p+1}}.
	\]
	Introducing the normalized amplitude $a = \sqrt{{2\Omega}/{\tilde{\gamma}}}\,\tilde{a}$, we obtain
	\[
	-\tilde{a}' 
	= \frac{1}{\sqrt{2}}
	\left(\frac{p + 1}{p}\right)^{\frac{1}{p+1}}
	\Omega^{\frac{2}{p+1} - \frac{1}{2}}
	\tilde{\gamma}^{\frac{1}{2} - \frac{1}{p+1}}
	\tilde{a}^{\frac{2}{p+1}} (1 - \tilde{a}^2)^{\frac{1}{p+1}}.
	\]
	Rearranging and integrating from the peak position $x_0=0$ (where $\tilde{a}(0)=1$) to $x$ yields
	\begin{align*}
		\frac{1}{2}&\!\int_1^{\tilde{a}^2(x)}\!
		s^{\left(\frac{1}{2} - \frac{1}{p+1} - 1\right)}
		(1 - s)^{\left(\frac{p}{p+1} - 1\right)}\,ds\\
		&= -\frac{1}{\sqrt{2}}
		\left(\frac{p + 1}{p}\right)^{\frac{1}{p+1}}
		\Omega^{\frac{2}{p+1} - \frac{1}{2}}
		\tilde{\gamma}^{\frac{1}{2} - \frac{1}{p+1}} x.
	\end{align*}
	This integral can be expressed in terms of the incomplete Beta function $B_z(a,b)$ \cite{abramowitz1965handbook}, leading to
	\begin{align*}
		B_{\tilde{a}^2(x)}&\!\left(\frac{1}{2} - \frac{1}{p+1}, \frac{p}{p+1}\right)
		- B\!\left(\frac{1}{2} - \frac{1}{p+1}, \frac{p}{p+1}\right)\\
		&= -\sqrt{2}
		\left(\frac{p + 1}{p}\right)^{\frac{1}{p+1}}
		\Omega^{\frac{2}{p+1} - \frac{1}{2}}
		\tilde{\gamma}^{\frac{1}{2} - \frac{1}{p+1}} x.
	\end{align*}
	In terms of the regularized incomplete Beta function
	\[
	I_z(a,b) = \frac{B_z(a,b)}{B(a,b)},
	\]
	this yields
	\begin{align*}
		&I_{\tilde{a}^2(x)}\!\left(\frac{1}{2} - \frac{1}{p+1}, \frac{p}{p+1}\right)\\
		&= 1
		-\frac{\sqrt{2}}{B\!\left(\frac{1}{2} - \frac{1}{p+1}, \frac{p}{p+1}\right)}
		\left(\frac{p + 1}{p}\right)^{\frac{1}{p+1}}
		\Omega^{\frac{2}{p+1} - \frac{1}{2}}
		\tilde{\gamma}^{\frac{1}{2} - \frac{1}{p+1}} x.
	\end{align*}
	Imposing symmetry and monotone decay, the soliton envelope can be written as
	\begin{equation}\label{sol_traj}
		a(x)
		= \sqrt{\frac{2\Omega}{\tilde{\gamma}}\,
			I^{-1}\!\left(
			1 -\alpha(p)|x|;
			\beta(p)\right)},
	\end{equation}
	where $I^{-1}$ denotes the inverse of the regularized incomplete Beta function with respect to its first argument \cite{abramowitz1965handbook}, and
	\begin{align*}
		&\alpha(p)=\frac{\sqrt{2}}{
			B\!\left(\frac{1}{2} - \frac{1}{p+1}, \frac{p}{p+1}\right)}
		\left(\frac{p + 1}{p}\right)^{\frac{1}{p+1}}
		\Omega^{\frac{2}{p+1} - \frac{1}{2}}
		\tilde{\gamma}^{\frac{1}{2} - \frac{1}{p+1}},\\
		&\beta(p)=\frac{1}{2} - \frac{1}{p+1}, \frac{p}{p+1}.
	\end{align*}

	\subsection{Numerical comparisons}
	
	\begin{figure}[tbhp!]
		\centering
		\includegraphics[width=0.65\linewidth]{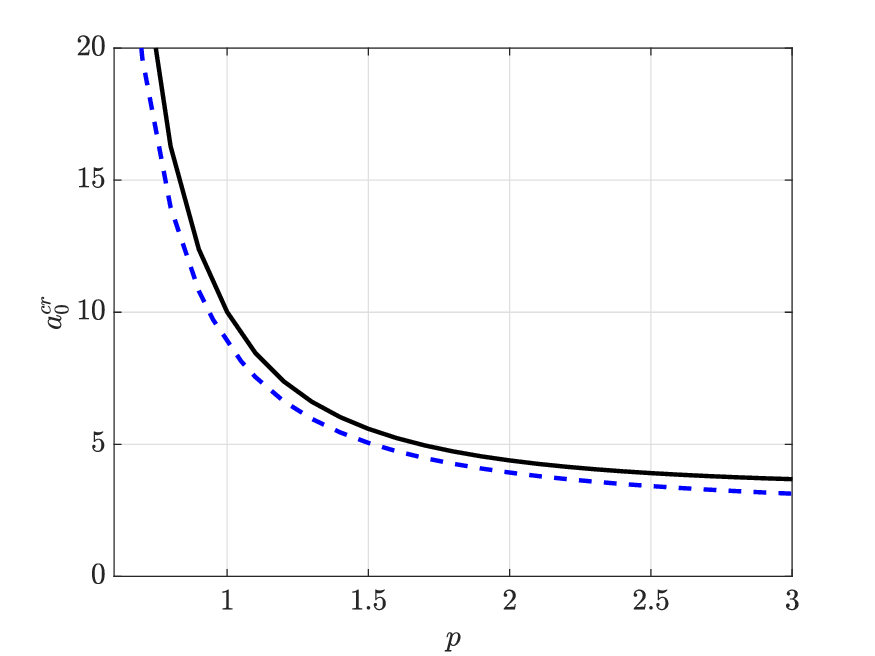}
		\caption{
			Critical driving amplitude $a_0^{\mathrm{cr}}$ versus coupling exponent $p$ for $\tilde{C}=0.1$, $\Omega=1$, and $\tilde{\gamma}=1/4$.  
			Black solid curve: numerical continuation.  
			Blue dashed curve: weak-coupling asymptotics \eqref{eq:approx}.  
			The approximation captures the overall trend.}
		\label{Fig4}
	\end{figure}

	Figure~\ref{Fig4} compares the numerically computed critical boundary amplitude $a_0^{\mathrm{cr}}$ with the analytical approximations derived in Sec.~\ref{sec6} for $\tilde{C}=0.1$.  
	The pseudo-arclength continuation results provide the reference curve.  
	The weak-coupling expansion \eqref{eq:approx} captures the qualitative dependence on $p$.
	
	In the strong-coupling regime, the amplitude-based criterion in Eq.~\eqref{strcamp} predicts that the supratransmission threshold should be independent of the nonlinearity exponent $p$.  
	However, the numerical results in Figs.~\ref{subfig:p_vs_Fcrit_c5} and \ref{Fig5} show otherwise. This deviation from the DpS prediction supports the conjecture that, for large coupling, supratransmission may be initiated by instabilities of the evanescent branch rather than by a fold bifurcation of stationary states, thereby introducing a residual $p$-dependence into the observed threshold.
	
	\section{Conclusion}\label{sec7}
	
	We have examined supratransmission in lattices with purely nonlinear coupling, a setting in which no linear pass band exists and energy transport is mediated entirely by nonlinear interactions.  
	Using an asymptotic reduction, the governing model was approximated by a discrete $p$-Schr\"odinger (DpS) equation, whose stationary and dynamical properties were analyzed in detail.  
	Analytical expressions for the critical driving amplitude were obtained in the weak- and strong-coupling regimes, revealing explicit dependence on the coupling exponent, driving frequency, and onsite nonlinearity.  
	Numerical continuation and direct simulations confirmed these predictions and showed that, for weak and intermediate coupling, supratransmission is associated with a fold (saddle-node) bifurcation of evanescent states in weak and intermediate coupling, whereas for strong coupling it appears to arise from a change in stability along the lower stationary branch.  
	These results extend previous studies based on linear dispersion and clarify how nonlinear coupling modifies the onset and mechanisms of energy transmission in discrete media.
	
	Several directions warrant further investigation. A first open question is whether the full governing equation exhibits instability-induced supratransmission in the strong-coupling regime corresponding to the onset of large energy propagation, as suggested by the DpS analysis; this requires a systematic continuation and stability study of evanescent waves in the original model. This transition appears qualitatively analogous to the second threshold $F_{c2}$ identified in \cite{maniadis2006energy}, whose precise dynamical origin likewise remains incompletely understood. The influence of boundary conditions, damping, and non-harmonic driving on the transmission threshold also remains to be quantified. From an analytical standpoint, higher-order reductions or multi-scale continuum limits may refine the accuracy of the DpS approximation near the threshold. Extensions to higher-dimensional lattices or to models with combined linear and nonlinear coupling could reveal additional transmission mechanisms. The insights obtained here may aid the design of tunable energy-guiding structures in mechanical metamaterials, nonlinear optical arrays, and micro-resonator networks.
	
	\section*{Acknowledgement}
	TYK acknowledged support from the Advanced Digital \& Additive Manufacturing (ADAM) Group (No.\ 8474000163). The authors thank the referees for their support, careful reading, and constructive comments. 
	
	\section*{Declaration of generative AI and AI-assisted technologies in the writing process}
	
	During the preparation of this work, DA and HS used ChatGPT and Grammarly to refine the language and improve sentence clarity throughout the manuscript. After using this tool/service, the authors reviewed and edited the content as needed and take full responsibility for the content of the publication.
	
	\bibliographystyle{unsrtnat}
	\bibliography{references}% Produces the bibliography via BibTeX.
	
\end{document}